# Li-Fi – Let There Be Light.

[1] Mihir Chauhan, [2] Aditya Kulai

[1][2]B. Tech IV Year EXTC, Electrical Department, VJTI, Mumbai, India.

***Abstract***: *The paper proclaims a summary of Li-Fi technology. It is an efficient data communication mechanism involving visible light as a medium of transmission. This monograph introduces the concept of Li-Fi and its working model. Furthermore, it discusses its benefits, shortcomings and examines ways to mitigate the drawbacks. Additionally, it analyzes its present applications and explores its future scope in the competitive wireless network market.*

## I. INTRODUCTION

Considering any situation, be it using wireless internet in public malls, contending for bandwidths at workshops and seminars or even engaging in home Wi-Fi networks, there is always a growing frustration when slow speeds of internet hampers and decelerates our work. This condition is engendered due to a substantial number of devices competing for a place in a comparatively smaller bandwidth of Radio Spectrum. An acclaimed German physicist, Dr Harald Haas came up with an effective solution to counter the ever-increasing problem. He introduced a much faster data transfer approach using visible light as a medium, which he termed as "Data through illumination". This is a wireless communication technique wherein visible light is used instead of radio waves. Various innovative aspects to optimize the communication have been initiated. This visible light communication method is extremely vital considering the ever-increasing population and hence the exponential rise in use of gadgets employing wireless networks. Hence, this innovation caters to the huge demand of a considerably narrow bandwidth of radio waves in the optical spectrum. There are various boons and even challenges to this methodology which will be precisely elucidated in the upcoming chapters of this paper.

## II. WORKING PRINCIPLE

The main ideology behind this technological innovation is that visible light illuminated by a light emitting diode (LED) is methodically amplitude modulated at the transmission end by rapid switching of LED lights at a speed not perceptible to human eye, whereas at the receiving end, photodiodes detect the modulated light and demodulates it to binary form by synchronized receiver circuits. In this way, data communication is successfully achieved.

## III. WORKING

The working of this technology is based on the type of working environment.

### A. Indoor Light Communication

Data sharing between several indoor devices such as a laptop, printer, television, desktop and an android phone can take place via visible light that illuminates the room. This type is characterised by an unlicensed spectrum which is highly dense, free and not affected by any Radio Frequency noise. Furthermore, it alleviates health problems.

### B. Cellular Communications

The axioms of wireless cellular communications can be extended to visible light communication to overcome the congestion of radio frequency spectrum in heterogeneous networks. The transmitters of optical cellular networks are termed as attocells. The optical cell helps mitigate the interference problems with the macro cellular network. These cells can be deployed with pico and femto cell environment which enables both free user mobility and high throughput of data. Such network can be an omnipresent wireless network structure.

The basic structure of visible light communication consists of a transmitter and a receiver. An independent transmitter and a receiver establish a downlink communication. The transmitter acts as a server and receiver as a client. Fig 1 shows communication between client and server. The server (transmitter) is interfaced with an array of LEDs and client (receiver) is interfaced to a photo diode. There can be multiple photo diodes and hence multiple clients. The distance between source LEDs and receiver photo diode is a function of the luminous intensity of LED arrangement, dark current, noise equivalent power and the responsivity of photodiode.

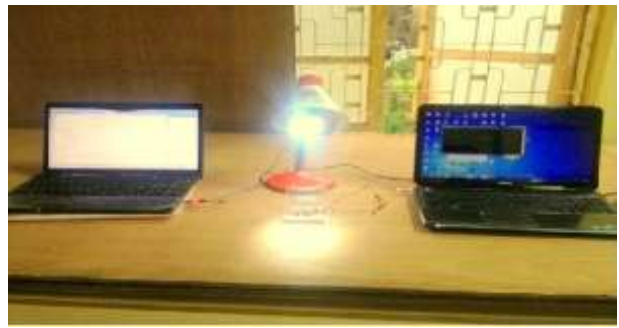

Fig. 1 Prototype of client-server communication using visible light as a medium of communication

A Li-Fi communication system and a full duplex communication is the necessity for both uplink and downlink communication. Therefore an integration of a transmitter and a receiver is required for the establishment of a full duplex communication.

### IV. ADVANTAGES

This data communication methodology has various advantages. They are explicitly presented below:

- Very high speeds of internet which can be more than 10 Gbps, which is way faster than other existing technologies like Wi-Fi, Bluetooth, IrDA, fire wire etc.
- Data is more secured and protected as the visible light is inherently unable to penetrate through walls unlike radio waves.
- Bandwidth of visible spectrum is 10,000 times more than that of radio spectrum, hence preventing any competitiveness for bandwidth.
- Implementation and maintenance costs are minimal compared to Wi-Fi.
- Elimination of health problems as visible light is the least harmful to humans in the optical spectrum.
- Enabling of Internet of Things on a large scale as more than 100 devices can be connected simultaneously.
- Easier and faster interpretation and conversion of data by the controller from binary stream of 1's and 0' to rapid LED switching.
- Possibility of data transmission through water, an edge over other transmission techniques.
- Efficiency of transmission of data is higher compared to other approaches
- Installation of Li-Fi can furnish wireless networks at locations where it is not feasible to establish connection through other technologies.

### V. CHALLENGES

However, there are some shortcomings in this technology which has been mentioned below:

- Communication is only possible if there is a direct line of sight between the transmitter and the receiver.
- Transmission is hindered and impaired when there is an opaque obstacle in the path of data transfer.
- Data communication only feasible inside closed rooms.
- Interference of other light sources such as sunlight can negate the efficiency of transmission.
- Communication is limited to point-to-point transfer when implemented at very high frequencies.

### VI. SOLUTION

The most vulnerable source of error is obstruction caused when an opaque obstacle is in between the source and the receiver. This obstruction is not sensed by the photo diode and thus the bits transferred during this interval of interference are lost. The photo diode receives a stream of zeroes during the period of hindrance, leading to misrepresentation of the data at the receiver. Fig 2 shows an original image transmitted by the transmitter and a distorted image received by the receiver due to obstruction to the path of light between transmitter and receiver.

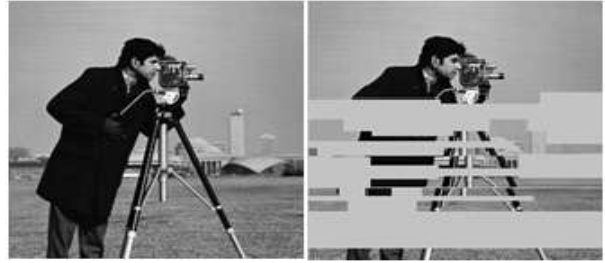

Fig. 2 Original transmitted image [Left], Distorted received image [Right]

Error due to obstruction can be eliminated by introducing a feedback sensor which helps sense obstruction. In this mechanism, whenever an obstruction is encountered, the data transmitting and receiving is paused. The data communication is resumed only after the obstruction has cleared the path of communication between the transmitter and receiver. In this way, the paramount shortcoming can be solved.

### VII. APPLICATIONS

The execution of this reliable technology is as follows:

- Excessive burden on RF spectrum can be reduced if some cellular network can utilize the visible spectrum.
- Can be used for underwater communication, especially for short range communication.
- Can be implemented in Hospitals as visible light do not interfere with any medical instrument unlike radio waves.
- It can be also used in smart lighting where street lamps can provide Li-Fi hotspots and switching of lighting can be controlled simultaneously.
- Can be used in petrochemical factories so as to avoid electromagnetic interference.
- Information exchange between different offices is enabled by point-to-point communication in Li-Fi technology.
- Traffic lights could provide information to the drivers via car GPS system enlightening them about accidents and congestion through pictures and videos.

- It initiates interaction with television as television shows can be downloaded as you watching. It avoids use of any phone applications for the same.
- Charging of electronic devices like phones, tablets and laptops if the devices incorporate a photoreceptor in the display screen.
- Safer internet in defense and national security organizations as the waves is impregnable through walls.

### VIII. CONCLUSION

This technology stands out with perfection, where data transfer is fast, efficient and significantly provides high level of security. Presently, Li-Fi technology is in its rudimentary stages. However with some developments and optimizations, the future scope of this reliable technology looks promising.

### ACKNOWLEDGEMENT

We are extremely grateful to the Centre of Excellence (CoE) department of V.J.T.I for providing us with all the essential resources. We thank our professors for elucidating all the necessary concepts related to wireless technology. We would also thank other members of our project group who helped to build a prototype of Li-Fi; wherein actual data such as pictures and videos were exchanged between two computer systems. We also appreciate the efforts taken by the students of our class for proof reading the paper and furnishing critical reviews for the same.

## References

[1] *Principles of LED light communication by* Prof. Herald Haas, Svilen Dimitrov.
[2] *Li-Fi (Light Fidelity)-The future technology in Wireless communication* - Jyoti Rani, Prerna Chauhan, Ritika Tripath. *(International Journal of Applied Engineering Research)*
[3] http://www.digplanet.com/wiki/Li-Fi
[4] http://www.lifi.com/pdfs/techbriefhowlifiworks.pdf
[5] http://en.wikipedia.org/wiki/Li-Fi
[6] *LED Lamp Based Visible Light Communication in Underwater Vehicle* - C.Periasamy , K.Vimal , D.Surender (International Journal of Engineering trends and technology)
[7] http://www.yuvaengineers.com/li-fi-technology-in-wireless-communication-revathi-ganesan/
[8] http://www.gizmag.com/li-fi-wireless-technology/32968/
[9] *Seminar on Li-Fi by* Tanshu.
[10] *Visible light communications: Challenges and possibilities by* O'brien, D.; Zeng, L.;Ha Le-Minh ;Faulkner, G. ;Walewski, J.W. ;Randel, S.
[11] *Visible Light Communiation* – Samsung Electronics, ETRI, VLCC and University of Oxford
[12] "High Efficient Li-Fi and Wi-Fi Technologies in Wireless Communication by Using Finch Protocol"- B.Narmada , P.Srinivasulu , P.Prasanna Murali Krishna .(International Journal of Engineering trends and technology)